\documentstyle[sprocl,epsfig]{article}

\def\Journal#1#2#3#4{{#1} {\bf #2}, #3 (#4)}

\def\aj{\em AJ}
\def\mnras{\em MNRAS}
\def\apj{\em ApJ}

\def\be{\begin{equation}}
\def\ee{\end{equation}}
\def\bea{\begin{eqnarray}}
\def\eea{\end{eqnarray}}

\begin{document}

\title{The All-Sky Asiago-ESO/RASS QSO Survey: Clustering and
Luminosity Function}

\author{A. Grazian}

\address{European Southern Observatory, Garching bei M\"unchen, Germany\\
E-mail: agrazian@eso.org}

\author{S. Cristiani}

\address{Trieste Astronomical Observatory, Trieste, Italy\\
E-mail: cristiani@ts.astro.it}

\author{A. Omizzolo}

\address{Astronomy Department, University of Padua,
Padua, Italy\\E-mail: omizzolo@pd.astro.it}


\maketitle\abstracts{ The {\bf AERQS} is a project aimed at the construction
of an all-sky statistically well-defined sample of local ($z\le 0.3$)
and bright ($R\le 15.5$) QSOs.
Present uncertainties on the theoretical modeling of the QSO evolution
are to be ascribed to the relatively small number of objects observed
in this magnitude and redshift domain.
We are filling this gap with a sample of 392 AGN covering all the sky at high
Galactic latitude: we
present here the Clustering analysis and the Luminosity Function of AGN at low
redshift. The {\bf AERQS} sets an important zero point about
the properties of the local QSO population
and the evolutionary pattern of QSOs between the present
epoch and the highest redshift.}

\section{Introduction}

The analysis of the Luminosity Function (LF) and Clustering
of QSOs is a fundamental cosmological tool to understand
their formation and evolution. QSOs, with normal galaxies, supply
detailed informations on the distribution of
Dark Matter Halos that are generally thought to constitute the ``tissue''
above which structures form and grow.
The lighting up of galaxies and QSOs depends
on how the baryons cool within the dark matter halos and form stars or
begin accreting onto the central BH, ending up as the only directly visible
peaks of a much larger, invisible structure. The detailed analysis of the
distribution of those peaks in the universe, i.e. their number and
clustering, can give important constraints on their evolution
and distinguish between the various favored models that explain the
formation of cosmic structures.
The theoretical interpretation of the QSO statistical properties
is not easy due to the number of ingredients used in theoretical models
and the degeneracies among them.
In particular there is not a unique combination of these parameters to
reproduce the LF and clustering of QSOs in the redshift range
$0.3\le z\le 2.2$. We have started a project,
the {\bf AERQS}
to find bright AGN in the local Universe and fill the existing gap
in present day large surveys:
it is paradoxical that in the era of 2QZ and SDSS, with
thousands of faint QSOs discovered even at high redshift (z=6.28),
still there are very few bright QSOs at low-z, compared to the
numbers at $z\sim 2$.
One of the main reasons is that
the surface density of low-z and bright QSOs is very low, of the order of
few times $10^{-2}$ per $deg^{2}$.
Another reason is that with optical information only is very difficult
to isolate efficiently bright QSOs from
billions of stars in large areas: a survey based on different selection
criteria is required.
We have used the X-ray emission, a key feature of the AGN population:
QSO candidates have been selected by cross-correlating the X-ray sources
in the RASS-BSC with {\em optically bright} objects in photographic plates.
The X-ray emission
is used only to compute an ``X-Optical color'' for the selection of AGN.
Therefore {\em the result of our selection cannot be
considered an identification of X-ray sources}.
The {\bf AERQS} is divided in three subsamples, two in the northern hemisphere
(GSC and USNO, described in Grazian et al. 2000[\cite{paper1}]) and the DSS
sample in the South (Grazian et al. 2002[\cite{paper2}]).
The spectroscopic identification of the candidates has been completed,
ending with 392 AGN (60$\%$ previously unknown)
over $\sim 14000$ $deg^2$ in the redshift
range $0.02\le z\le 2.04$. The redshift distribution shows a peak around
$z\sim 0.1$ with an extended tail up to $z=0.4$. Five AGN with
$0.6\le z\le 2.04$ are possibly magnified/lensed objects.

\section{The Luminosity Function}

\begin{figure}
\vskip -2.5 true cm
\centerline{\epsfig{figure=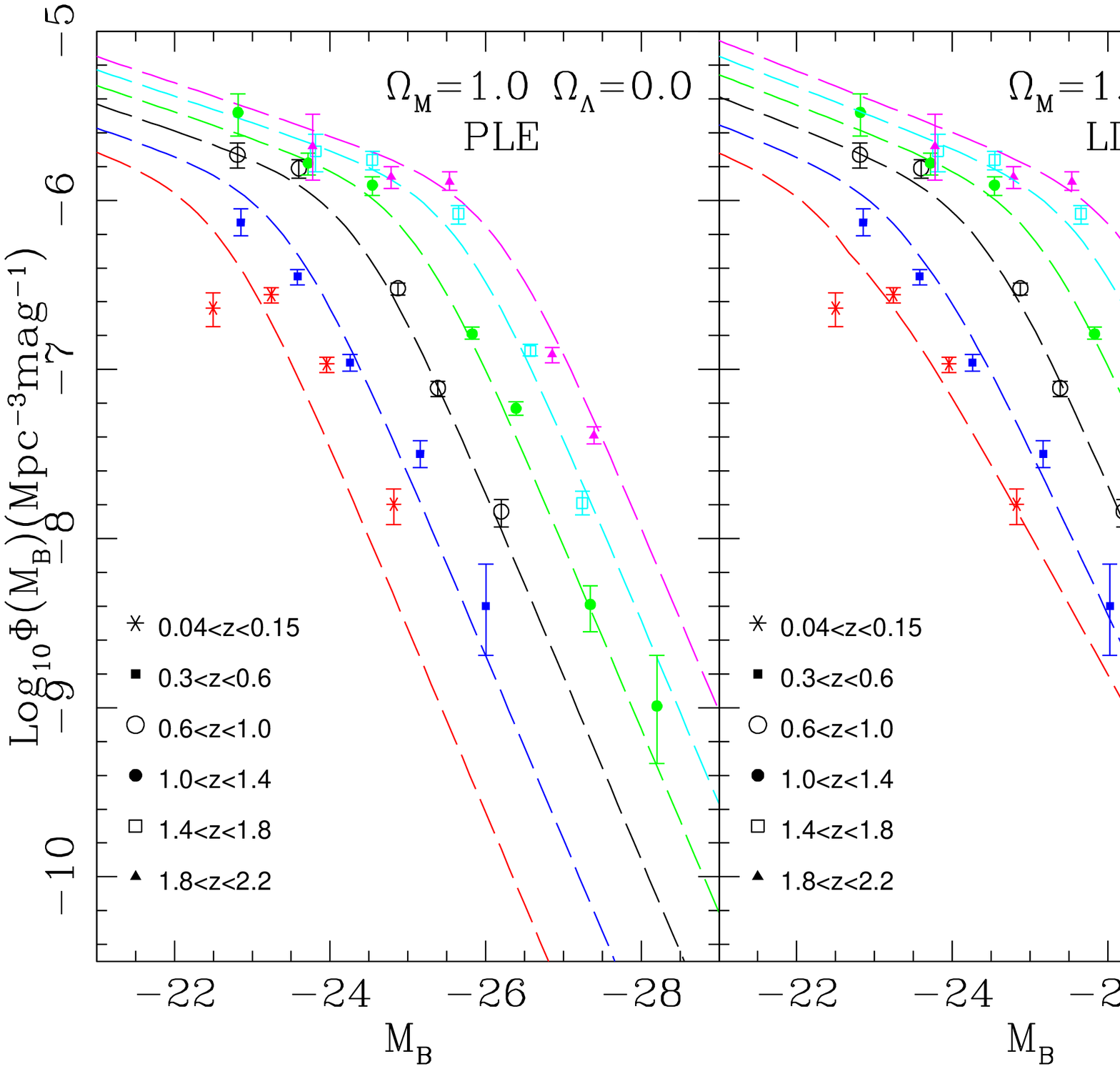, width=9cm, height=8cm}}
\vskip -0.3 true cm
\caption{
The LF of QSOs compared with a parameterization of PLE and
LDLE. The points in the range $0.04\le z \le 0.15$ are the result of the
{\bf AERQS}, the remaining data are derived from [3].}
\label{lf}
\end{figure}

The general behavior of the QSO Optical Luminosity Function (OLF)
is well established in the redshift
interval $0.3<z<2.2$ for which color techniques provide reliable selection
criteria.
A pure luminosity evolution (PLE) appears to reasonably describe the data
in the interval $0.6<z<2.2$.  
In the range $0.3<z<0.6$ the OLF appears to be flatter than
at higher redshifts, requiring a luminosity dependent
luminosity evolution (LDLE[\cite{lfc97}]).
To provide a statistically solid basis for the LDLE
pattern and to investigate whether such a trend
persists (and possibly becomes more evident)
we decided to investigate the OLF at $z\le 0.3$.
To compute the LF at $0.04\le z\le 0.15$ we have used the generalized
$1/V_{max}$ ``coherent'' estimator in a slightly modified
version which tries to estimate in an unbiased manner the volume-luminosity
space ``available'' to each object [\cite{pc00}]
and takes into account the evolution of the LF within the redshift
intervals.
Fig. \ref{lf} shows that an LDLE parameterization can reproduce the data in a
much more satisfactory way than PLE.
This behavior provides an interesting clue for the
physical interpretation of the QSO evolving population
[\cite{CV02,KH00,paper1}].
\section{The Clustering Evolution}

\begin{figure}
\vskip -2.5 true cm
\centerline{\epsfig{figure=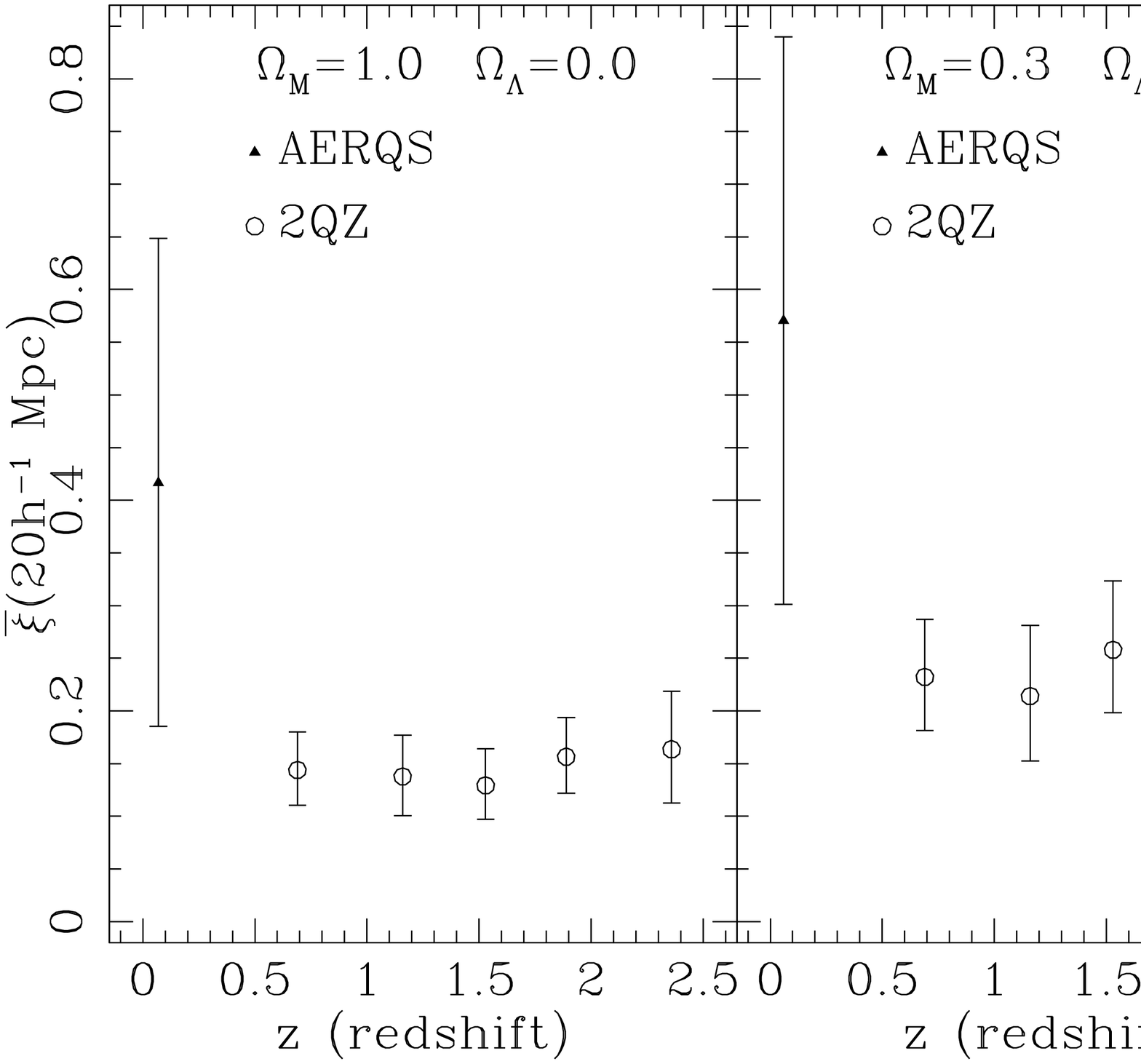, width=9cm, height=8cm}}
\vskip -0.3 true cm
\caption{
The clustering evolution of QSOs with redshift for two different cosmologies.
Open circles are from 2QZ, filled triangle shows the {\bf AERQS}~result.}
\label{clust}
\end{figure}

Recently Croom[\cite{croom}], using more than $10^4$ QSOs taken from the
preliminary data release of 2QZ, measured the redshift evolution of
QSO clustering.
For an Einstein-de Sitter Universe there is no
significant evolution in comoving
coordinates over the redshift range $0.3\le z\le 2.9$,
whereas for $\Omega_{M}=0.3$
and $\Omega_{\Lambda}=0.7$ clustering shows a
marginal increase at high-z.
The clustering analysis of QSOs in the {\bf AERQS} at $z\le 0.3$ has been
carried out
simply by computing the two point correlation function (TPCF)
in the redshift space.
We have used the minimum variance estimator suggested by Landy \&
Szalay[\cite{ls93}] to calculate $\xi(r)$.
Fig. \ref{clust} compares the evolution of the TPCF
$\xi$ at $0.3\le z\le 2.9$ from 2QZ with the
value at $z\le 0.3$ from {\bf AERQS}.
The redshift evolution shows a marginal increase of $\xi$ at high-z
and an increase at low-z. Such a behavior is generally
predicted by Bagla[\cite{bagla}]: at high-z bright
objects reside in Massive DMHs, strongly biased and consequently
significantly clustered. The bias of the DMHs decreases with decreasing
redshift, but the clustering of the underlying matter increases. At low-z
AGN are tracing highly clustered but less biased DMHs.
\section{Conclusion}

We have built a well defined sample
of {\em 392 optically bright ($R\le 15.5$) AGN (60$\%$ are new identifications)
in the redshift range $0.02\le z\le 0.30$}, filling the gap
between the local Universe and the highest redshifts.
The {\em QSO Optical LF}
at $0.04 \le z \le 0.15$ is computed and shown to be consistent with a
LDLE Evolution of the type derived by La Franca \& Cristiani[\cite{lfc97}]
in the redshift range $0.3 \le z \le 2.2$.
The {\em TPCF of QSOs} is constant or mildly increasing at high-z
and increasing towards low-z.
The QSO LF, together with the
correlation function, at $z\le 0.3$ constitutes a
{\em fundamental zero point} for a detailed modeling of the formation and
evolution of cosmic structures in the Universe.


\section*{References}
\begin{thebibliography}{99}
\bibitem{paper1} Grazian, A. et al. \Journal{\aj}{119}{2540}{2000}.

\bibitem{paper2} Grazian, A. et al. 2002, astro-ph/0208134.

\bibitem{lfc97} La Franca, F. \& Cristiani, S. \Journal{\aj}{113}{1517}{1997}.

\bibitem{pc00} Page, M. J. \& Carrera, F. J. \Journal{\mnras}{311}{433}{2000}.

\bibitem{CV02} Cavaliere, A. \& Vittorini, V. \Journal{\apj}{570}{114}{2002}.

\bibitem{KH00} Kauffmann, G. \& Haehnelt, M. \Journal{\mnras}{311}{576}{2000}.

\bibitem{croom} Croom, S. M. et al. \Journal{\mnras}{325}{483}{2001}.

\bibitem{ls93} Landy, S. D. \& Szalay, A. S. \Journal{\apj}{412}{64}{1993}.

\bibitem{bagla} Bagla, J. S. \Journal{\mnras}{299}{417}{1998}.

\end{thebibliography}
\end{document}